\documentclass[twocolumn]{IEEEtran}
\usepackage{caption}
\usepackage{subcaption}
\usepackage{lineno,hyperref}
\usepackage{comment}
\usepackage{amsmath,amsfonts,amsthm,bm}
\usepackage{graphicx}
\usepackage{longtable,tabularx}
\usepackage{booktabs}

\usepackage{algorithm}
\usepackage{algpseudocode}
\newcommand{\spacedhrule}{\vspace{1mm}\hrule\vspace{1mm}}

\newcommand{\SCUC}{\text{SCUC}}
\newcommand{\SCED}{\text{RT-SCED}}

\modulolinenumbers[5]
\usepackage[margin=0.67 in]{geometry}
\usepackage{geometry}
\usepackage[numbers]{natbib}
\usepackage{setspace}
\usepackage{soul}
\usepackage{lineno}
\usepackage[usenames,dvipsnames]{xcolor}
\usepackage{url}

\title{Just-In-Time Learning for Operational Risk Assessment in Power Grids}

\author{Oliver Stover \IEEEauthorrefmark{1}, Pranav Karve \IEEEauthorrefmark{1}, Sankaran Mahadevan \IEEEauthorrefmark{2} \\ Wenbo Chen \IEEEauthorrefmark{3}, Haoruo Zhao \IEEEauthorrefmark{3}, Mathieu Tanneau \IEEEauthorrefmark{3}, Pascal Van Hentenryck \IEEEauthorrefmark{3}
\thanks{
\IEEEauthorblockA{\IEEEauthorrefmark{1} \IEEEauthorrefmark{2} Vanderbilt University, Nashville, TN 37235, USA.}\\
\IEEEauthorblockA{\IEEEauthorrefmark{3} Georgia Institute of Technology, Atlanta, GA 30332 USA.}\\
\IEEEauthorblockA{\IEEEauthorrefmark{1} These two authors made equal contributions to this paper.} \\
\IEEEauthorblockA{\IEEEauthorrefmark{2} Corresponding author, E-mail: sankaran.mahadevan@vanderbilt.edu}
}}

\begin{document}
\maketitle
\begin{abstract}
In a grid with a significant share of renewable generation, operators will need additional tools to evaluate the operational risk due to the increased volatility in load and generation. The computational requirements of the forward uncertainty propagation problem, which must solve numerous security-constrained economic dispatch (SCED) optimizations, is a major barrier for such real-time risk assessment. This paper proposes a Just-In-Time Risk Assessment Learning Framework (JITRALF) as an alternative. JITRALF trains \emph{risk surrogates}, one for each hour in the day, using Machine Learning (ML) to predict the quantities needed to estimate risk, without explicitly solving the SCED problem. This significantly reduces the computational burden of the forward uncertainty propagation and allows for fast, real-time risk estimation. The paper also
proposes a novel, asymmetric loss function and shows that models trained using the asymmetric loss perform better than those using symmetric loss functions. JITRALF is evaluated on the French transmission system for assessing the risk of insufficient operating reserves, the risk of load shedding, and the expected operating cost.
\end{abstract}

\section{Introduction}
\label{sec:intro}

American and global bulk electrical systems (BES) are projected to have significant shares of renewable generation in the future~\cite{zhongming2021nuclear,ren,nercLongTerm}. This new generation mix
 introduces new challenges due to increased volatility in load and generation
 and more significant prediction errors. Currently, grid operators manage uncertainty and risk in daily operations by acquiring excess unused capacity (known as reserves) and leveraging their experience. However, these implicit risk management approaches may no longer be sufficient to effectively manage risk in operational regimes with higher uncertainty and lower safety margins. Consequently, explicit methods to assess operational risk are needed to inform grid operators of the risks associated with an operational decision and to enable proactive risk management~\cite{risk_metrics1}.

Typically, the explicit risk assessment framework propagates Monte Carlo (MC) samples of stochastic input variables (wind/solar generation and load) through the system operational model to obtain samples of the (output) quantities of interest (QoIs). This \emph{forward propagation} of uncertainty, in the case of power systems, requires the solution of numerous Security-Constrained Economic Dispatch (SCED) optimization problems. The samples of QoIs are then used to compute conditional expectations, probabilities of failure, and (monetary) risk corresponding to relevant failure modes~\cite{risk_metrics1}. The samples of the stochastic input variables are obtained using probabilistic forecasts of those variables. These forecasts are updated throughout the day using the most up-to-date information. Real-time system risk assessment could also be repeated, using this updated forecast, to obtain the most up-to-date information about the system risk profile. With real-time risk assessment, a grid operator may be able to accept day-ahead generator schedules with higher risk, but lower cost~\cite{risk_metrics1}, and delay the risk mitigating decisions (e.g., committing additional generators) until the high-risk event becomes credible. Real-time risk estimation could also provide the most up-to-date estimate of the magnitude of additional resources required to mitigate the risk. Since grid operators will only commit the additional generators if the real-time risk profile predicts upcoming adverse conditions and will only commit the smallest amount of capacity needed to mitigate the risk, they will be able to operate the grid at lower cost, while still maintaining acceptable levels of system reliability/risk. In this manner, the ability to delay commitment decisions and minimize the magnitude of interventions, while maintaining grid reliability, has significant cost-saving potential. {\em Unfortunately, real-time risk assessment
using Monte-Carlo simulation and the forward propagation of uncertainty is too expensive computationally to provide updated risk estimates in real time.}

To address this computational bottleneck, this paper proposes a Just-In-Time Risk Assessment Learning Framework (JITRALF).  JITRALF trains risk surrogates to predict the QoIs directly, without explicitly solving numerous SCED optimizations. Moreover, to account for the fact that generator commitments
change hourly, which complicates the learning task substantially, JITRALF uses the \emph{just-in-time} learning paradigm introduced in \cite{Chen2022_JustInTimeSCED}, wherein machine-learning models are trained after the execution of the DA-FRAC, and training data consists of the simulations executed in the DA risk-assessment. JITRALF also includes a novel, asymmetric loss function to capture the greater importance of prediction accuracy in unsafe regions, i.e., input regions where the QoIs are near or beyond
acceptable operating limits.  JITRALF has been evaluated on the French transmission system for assessing the risk of insufficient operating reserves,
the risk of load shedding, and the expected operating cost. Experimental results show that JITRALF achieves good prediction accuracy for all three
cases, and correctly identifies risky operating conditions on high-risk scenarios. Moreover, the experiments also show that machine-learning models trained using the asymmetric loss perform better than those using symmetric
loss functions.

The contributions of JITRALF can be summarized as follows:
\begin{enumerate}
    \item JITRALF uses a Just-In-Time Risk Assessment Learning Framework that leverages risk surrogates to overcome the computational barriers faced by traditional risk assessment approaches.
    
    \item JITRALF is a just-time learning methodology that trains dedicated risk surrogates for each hour of the day, addressing the fact that generator commitments change hourly.
    
    \item JITRALF leverages probabilistic forecasting, digital twins of ISO operations, and principled risk/reliability metrics in order to obtain a high-fidelity risk assessment framework.
    
    \item JITRALF includes a novel, asymmetric loss function to capture the greater importance of prediction accuracy in unsafe regions.
    
    \item The potential benefits of JITRALF have been demonstrated on the French transmission system for assessing the risk of insufficient operating reserves, the risk of load shedding, and the expected operating cost.
\end{enumerate}
\noindent
The rest of the paper is organized as follows. Section \ref{sec:relatedWork} briefly discusses related work and positions the paper. 
Section \ref{sec:risk_assessment:MISO_operations} describes Midcontinent Independent System Operator's, or MISO's, current operations. The authors' previously proposed risk-assessment framework is summarized in Section~\ref{sec:risk}.
The proposed just-in-time learning approach and the framework to train and validate the ML surrogate models are developed in Section~\ref{sec:MLrisk}.
A synthetic version of the French RTE grid, used to illustrate operational risk estimation, is described in Section~\ref{sec:numex}.
Risk learning and estimation results for this synthetic grid are shown in Section~\ref{sec:res}.
Conclusions and future research areas are discussed in Section~\ref{sec:conclusion}.

\section{Related Work}
\label{sec:relatedWork}

The use of computationally inexpensive surrogate models for easing the computational burden of probabilistic analyses is not new. In many engineering problems, surrogates for high-fidelity physics models are trained to enable and/or expedite MC simulations~\cite{vohra2020fast,sudret2017surrogate}. Neural networks~\cite{papadrakakis2002reliability}, support vector machines~\cite{rocco2002fast}, Gaussian process models~\cite{bichon2008efficient}, polynomial chaos expansion (PCE) models~\cite{jakeman2022surrogate}, and many other surrogate model forms have been successfully used for this purpose. The surrogate models are usually developed to replace computationally demanding system models. JITRALF builds surrogate models to replace the time-consuming, operational optimization for power grids. This is a challenging task because the surrogate model needs to predict QoIs corresponding to the given sample of stochastic inputs without explicitly considering the objective (cost) function, the decision variables, and the physical constraints on the system. The samples of QoIs generated by repeated execution of the computationally inexpensive surrogate model are then used for reliability/risk assessment. 

Various machine learning models have been developed to enable fast prediction of adverse events in the power systems domain. These include ML models for the prediction of grid stability~\cite{nauck2021predicting,sun2022fast,yang2021power}, outages due to extreme weather~\cite{haseltine2017prediction,eskandarpour2016machine}, and power quality disturbances~\cite{khokhar2015comprehensive,alimi2020review}. However, these models have not been deployed to support operational decision-making in commercial power grids. They have been developed for very small, synthetic grids. Additionally, they do not consider the day-to-day changes in the grid topology due to different unit commitments and transmission line outages. Even for a small-sized, real-world power grid, generating training data for all possible grid topologies (i.e., all possible combinations of operating statuses of generators and transmission lines) and weather conditions is not possible, because of the combinatorial explosion of possible operating states. In order to tackle this impediment, JITRALF uses the just-in-time learning paradigm of ~\cite{Chen2022_JustInTimeSCED} to train risk surrogate models each day and for each hour of the day based on the \emph{day-ahead} weather forecasts, generator commitments, and grid topologies. 



\section{The MISO Operational Pipeline}
\label{sec:risk_assessment:MISO_operations}

\begin{figure}[!t]
    \centering
    \includegraphics[width=\columnwidth]{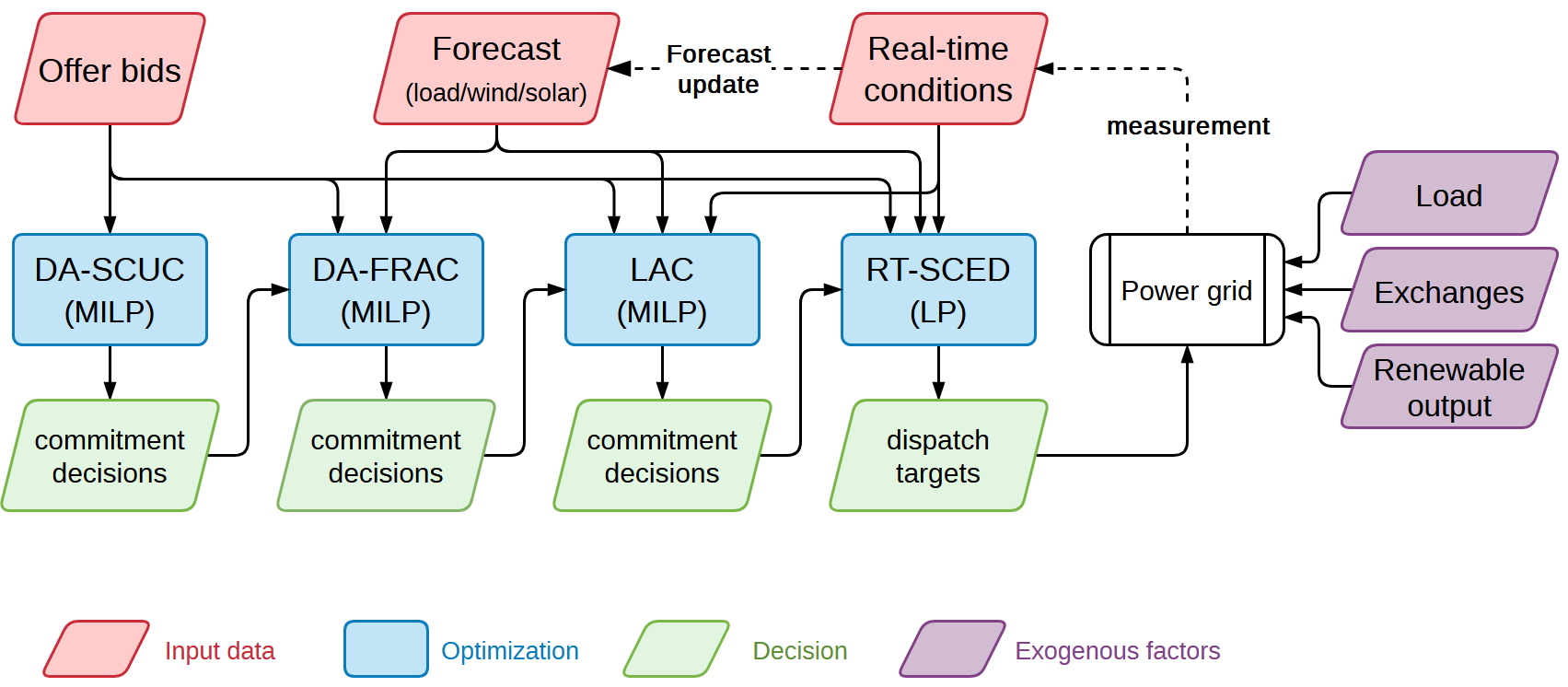}
    \caption{The MISO Optimization Pipeline.}
    \label{fig:optimization:miso_pipeline}
\end{figure}
    
This section describes the operating pipeline of MISO, the Mid-Continent Independent System Operator.
The pipeline, depicted in Figure \ref{fig:optimization:miso_pipeline}, covers the period going from the day-ahead market clearing to real-time operations.
The Day-Ahead (DA) energy and reserve market clearing takes place between 10am and 1pm prior to the operating day.
The DA market is a Security-Constrained Unit Commitment (SCUC) that takes as inputs supply and demand bids  from each market participant, and produces commitment decisions for each generator.
A Forward Reliability Assessment Commitment (DA-FRAC) is executed after the DA market clearing.
The DA-FRAC takes, as inputs, the previous commitment decisions and current forecasts for load and renewable generators, and may commit additional generation units to serve the predicted demand reliably. The role of the FRAC is to address the discrepancies between cleared and predicted net demand, by committing additional units as needed.
Throughout the operating day, a Look-Ahead Commitment (LAC) optimization is executed every 15 minutes with a 3-hour look-ahead window.
Again, the LAC may commit additional units to accommodate previously unforeseen changes in net load.
In addition, every 5 minutes, MISO clears the real-time market by solving a Security-Constrained Economic Dispatch problem (RT-SCED), which computes active power dispatch decisions for all online generators.
These dispatch decisions are sent to (human) operators who, after performing additional AC power flow studies, may adjust dispatch targets as appropriate.

The DA-SCUC, DA-FRAC, and LAC are security-constrained unit-commitment (SCUC) problems and are modeled as mixed-integer linear programming (MILP) problems of the form
\begin{subequations}
\begin{align}
    \label{eq:SCUC:obj}
    (\SCUC) \quad \min_{x, y} \quad & c(x, y)\\
    s.t. \quad
    \label{eq:SCUC:con:linear}
    & Ax + By \geq b,\\
    \label{eq:SCUC:con:discrete}
    & x \in \mathcal{X},
\end{align}
\end{subequations}
where $x$ denote the (discrete) commitment and startup/shutdown variables, and $y$ denotes the (continuous) energy and reserve dispatch variables, respectively.
The objective $c(x, y)$ minimizes production costs and constraint violation penalties.
Constraints \eqref{eq:SCUC:con:linear} include minimum/maximum limits, ramping constraints, thermal constraints, power balance and reserve requirements, and constraints \eqref{eq:SCUC:con:discrete} denote commitment-related constraints such as minimum up and down times for each generator.
The full formulations of DA-SCUC, DA-FRAC and LAC are described in \cite{BPM_002}.
The RT-SCED is a linear programming (LP) problem of the form
\begin{subequations}
\begin{align}
    (\SCED(z)) \quad \min_{y} \quad & q_{z}^{T}y\\
    s.t. \quad
    & H_{z} y \geq h_{z},
\end{align}
\end{subequations}
where $z$ denotes input parameters such as current load, renewable production, generators' online status and previous output, etc., and $y$ denotes the dispatch variables.
MISO's RT-SCED is a single-period, deterministic problem, which uses a DC approximation with linearized line losses \cite{BPM_002,Ma2009_MISOSCED}.

The DA-SCUC, DA-FRAC and LAC only output commitment decisions; energy and reserve dispatch decisions are not final until the RT-SCED is executed.
The DA-SCUC, DA-FRAC, LAC and RT-SCED problems must be solved within 20, 40, 10 and 2 minutes, respectively.
Despite significant progress in the resolution of UC problems \cite{Anjos2017_UCinElectricEnergySystems,Knueven2019_MILPFormulationsforSCUC,chen2022security}, industrial cases remain computationally and numerically challenging to solve \cite{chen2022security,Chen2021_HIPPO}.

\section{Operational Risk Assessment for Power Grids}
\label{sec:risk}

\begin{figure}[!ht]
    \centering
    \includegraphics[width=0.9\columnwidth]{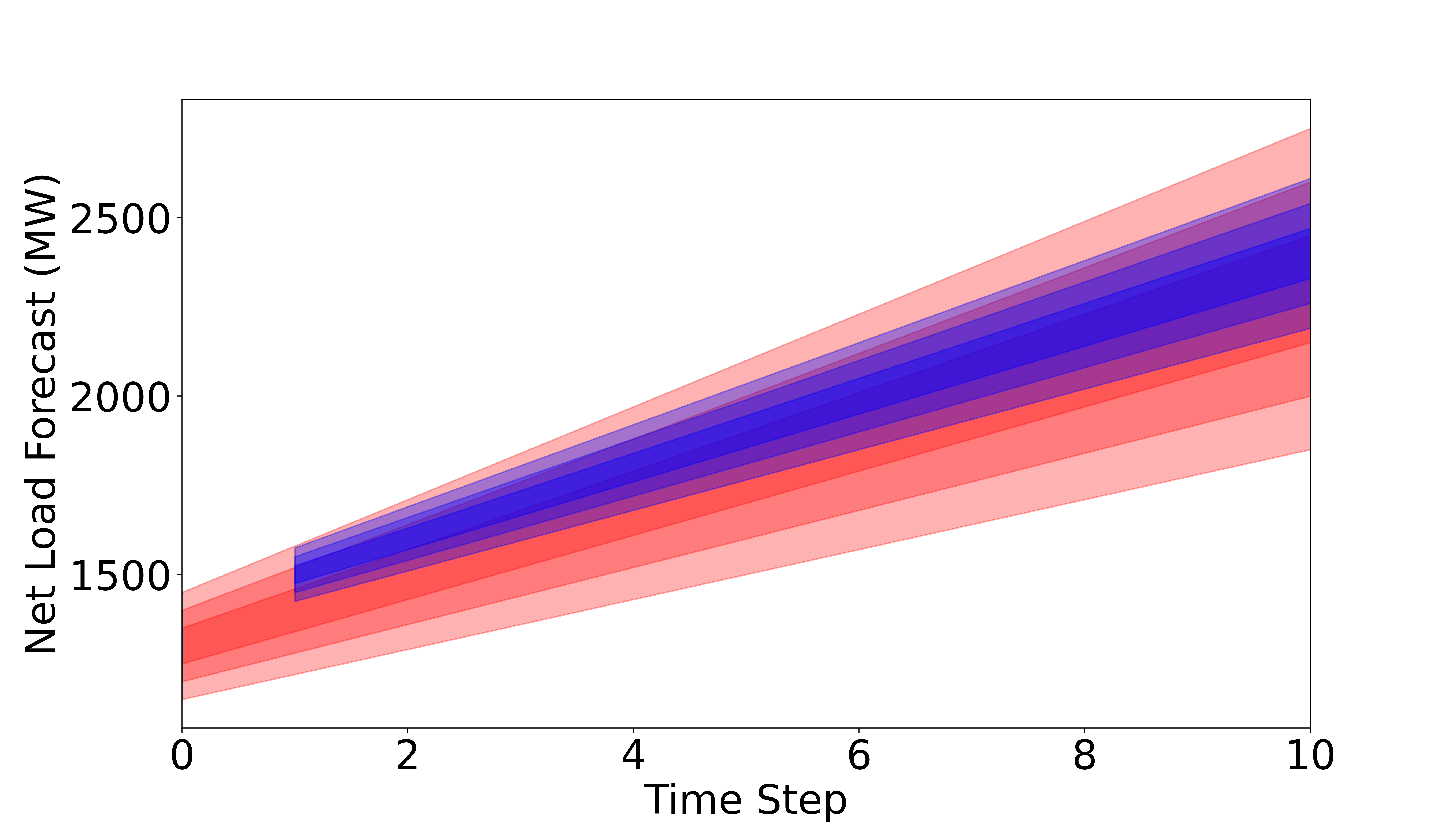}
    \caption{%
    Illustration of the net load probabilistic forecast update.
    \textcolor{red}{\textbf{Red}}: initial probabilistic forecast for net load, produced at $t{=}0$.
    \textcolor{blue}{\textbf{Blue}}: updated probabilistic forecast, produced at $t{=}1$.
    Darker shades indicate higher likelihood.}
    \label{fig:forecast}
\end{figure}

The present work leverages the risk and reliability assessment framework proposed in Stover et al \cite{risk_metrics1}, to perform both day-ahead (DA) and real-time (RT) operational risk assessments.
The day-ahead assessment follows the execution of the DA-FRAC and evaluates operational risk for the next day. This allows operators to take preemptive actions or adjust an overly-conservative portfolio.
The real-time assessment is executed throughout the operating day, and evaluates the operational risk for the next hour, given the most up-to-date system conditions.
Indeed, as the day progresses, operators receive updated forecasts, as illustrated in Figure \ref{fig:forecast}: in the figure, at $t=1$, higher net load values become more likely, and the initial risk profile is no longer valid and must be updated.

\begin{algorithm}[!ht]
    \caption{Sampling-based Risk Assessment}
    \label{alg:MC_Risk_Assessment}
    \begin{algorithmic}[1]
        \Statex \textbf{Inputs:}
            \begin{itemize}
                \item[-] Simulation horizon $\mathcal{T} = \{1, ..., T\}$
                \item[-] Generators' commitments $\mathbf{x} = (x_{1}, ..., x_{T})$
                \item[-] Generators' initial output $y_{0}$
                \item[-] MC samples $\mathbf{\xi}^{i} = (\xi^{i}_{1}, ..., \xi^{i}_{T}), \forall i \in \{1, ..., N\}$
                \item[-] QoI function $Q$
                \item[-] Risk/reliability computation function $R$
            \end{itemize}
        \Statex \textbf{Outputs:} Risk/reliability estimate $\mathfrak{R}$
        \spacedhrule
        \For{$i = 1, ..., N$} \label{alg:MC_risk:MC_begin}
            \State $\mathbf{y}^{i} = ()$ \label{alg:MC_risk:MC:init}
            \For{$t=1, ..., T$}  \Comment{MC simulation} \label{alg:MC_risk:MC:For}
                \State $z^{i}_{t} = (x_{t}, y^{i}_{t-1}, \xi^{i}_{t})$
                \State $y^{i}_{t} \gets \text{RTSCED}(z^{i}_{t})$
                    \Comment{Solve RT-SCED} \label{alg:MC_risk:MC:RTSCED}
                \State $\mathbf{y}^{i} \gets (\mathbf{y}^{i}, y^{i}_{t})$
                    \Comment{Record dispatches} \label{alg:MC_risk:MC:save}
            \EndFor \label{alg:MC_risk:MC:EndFor}
            \State $\mathbf{q}^{i} \gets Q(\mathbf{x}^{i}, \mathbf{y}^{i}, \mathbf{\xi}^{i})$
                \Comment{QoI evaluation} \label{alg:MC_risk:MC:QoI}
        \EndFor \label{alg:MC_risk:MC_end}
        \State $\mathfrak{R} = R(\mathbf{q}^{1}, ..., \mathbf{q}^{N})$
            \Comment{Risk/reliability computation} \label{alg:MC_risk:risk}
        \State \Return $\mathfrak{R}$ \label{alg:MC_risk:return}
    \end{algorithmic}
\end{algorithm}

The risk-assessment methodology employs the MC sampling-based approach formalized in Algorithm \ref{alg:MC_Risk_Assessment}.
Each MC sample $\xi^{i} = (\xi^{i}_{1}, ..., \xi^{i}_{T})$ is a possible realization of the uncertain variables, such as load and renewable production, over the considered simulation horizon.
Those samples are obtained from a probabilistic forecasting model that captures the joint distribution of these variables, e.g., a DeepAR \cite{Salinas2020_DeepAR} or Temporal Fusion Transformer \cite{Lim2021_TFT} model.
For ease of notation, all MC samples are assumed to have equal probability; this assumption is easily relaxed \cite{risk_metrics1}.
Given the current system condition and dispatch solution, the function $Q$ computes the value of a QoI, e.g., the total operating cost, system-wide reserve capacity, or system-wide ramping capacity, which can then be used to compute the probability and risk of failure events \cite{risk_metrics1}.
The risk computation function $R$ will be described shortly.

Algorithm \ref{alg:MC_Risk_Assessment} then proceeds as follows.
First, for each MC sample $i \in \{1, ..., N\}$, the system's behavior is simulated over the horizon $\mathcal{T}$.
This is done by solving, at lines \ref{alg:MC_risk:MC:For}--\ref{alg:MC_risk:MC:EndFor}, a sequence of $T$ RT-SCED optimizations -- one every 5 minutes, and recording the corresponding solutions.
Then, in line \ref{alg:MC_risk:MC:QoI}, the resulting $T$ solutions are used to compute $\mathbf{q}^{i}$, the QoI value under scenario $i$.
Note that lines \ref{alg:MC_risk:MC_begin}--\ref{alg:MC_risk:MC_end} are independent MC simulations, and can therefore be executed in parallel.
In contrast, lines \ref{alg:MC_risk:MC:For}--\ref{alg:MC_risk:MC:EndFor} must be executed sequentially, because each RTSCED problem uses the previous solution as input.
Finally, in line \ref{alg:MC_risk:risk}, the $N$ QoI samples $\mathbf{q}^{1}, ..., \mathbf{q}^{N}$ are combined into a risk estimate $\mathfrak{R}$, which is returned in line \ref{alg:MC_risk:return}.

Stover et al. \cite{risk_metrics1} define three levels of risk and reliability metrics, which correspond to different choices for the risk/reliability computation function $R$.
Denote by $q$ the value of a given QoI, e.g., the total system reserves, and by $\bar{q}$ the (lower) limit of acceptable values for the QoI.
For instance, $\bar{q}$ may denote the minimum reserve requirement set by system operators.
Level-1 metrics measure the conditional expectation of the QoI in the worst $\alpha\%$ cases, 
\begin{align}
    \label{eq:risk:level1}
    \forall t \in \mathcal{T}, R_{\alpha}(q^{1}_{t}, ..., q^{N}_{t}) &= \frac{100}{\alpha N} \sum_{i \, | \, q^{i}_{t} \leq q^{(\alpha)}_{t}} q^{i}_{t},
\end{align}
where $q^{(\alpha)}_{t}$ denotes the $\alpha$-th percentile of $\{q^{1}_{t}, ..., q^{N}_{t}\}$.
Level-2 metrics measure reliability by estimating the probability of an adverse event, e.g., the QoI violating the safety limit.
This yields
\begin{align}
    \label{eq:risk:level2}
    \forall t \in \mathcal{T}, R(q^{1}_{t}, ..., q^{N}_{t}) &= \frac{1}{N} \sum_{i \, | \, q^{i}_{t} < \bar{q}} 1.
\end{align}
Finally, Level-3 metrics measure the financial consequences of adverse events.
Here, each MC sample provides a different consequence cost $C(\bar{q}, q^{i}_{t})$ for violating the safety limit.
The MC simulation process thus provides a probability distribution of the risk.
This work employs the mean value of the risk distribution as a risk metric, i.e.,
\begin{align}
    \label{eq:risk:level3}
    \forall t \in \mathcal{T}, R(q^{1}_{t}, ..., q^{N}_{t}) &= \frac{1}{N} \sum_{i} C(\bar{q}, q^{i}_{t}).
\end{align}
For instance, in MISO, the consequence cost of failing to meet minimum reserve requirements is defined via reserve demand curves \cite{MISO_Tariff28}.
Note that Equations \eqref{eq:risk:level1}--\eqref{eq:risk:level3} can easily be adapted for risk/reliability computation involving a variety of QoIs, e.g., reserve capacity, load shed, transmission violations, etc.
The reader is referred to \cite{risk_metrics1} for a detailed discussion on the quantification of the risk and reliability.

Each execution of Algorithm \ref{alg:MC_Risk_Assessment} requires in the order of a thousand MC samples to obtain an accurate risk estimate.
This represents $288,000$ and $12,000$ RT-SCED optimizations per day-ahead and real-time risk assessment, respectively.
While individual MC simulations can be carried out in parallel, optimization-based risk assessment requires significant computational resources.
Furthermore, as the complexity of RT-SCED problems increases in the future, the present methodology may not be tractable for real-time risk assessment.
The paper addresses this computational barrier via the use of machine-learning techniques, which are described next.

\section{Just-In-Time Risk Assessment Learning}
\label{sec:MLrisk}

\begin{figure}[!ht]
    \centering
    \includegraphics[width=0.45\textwidth]{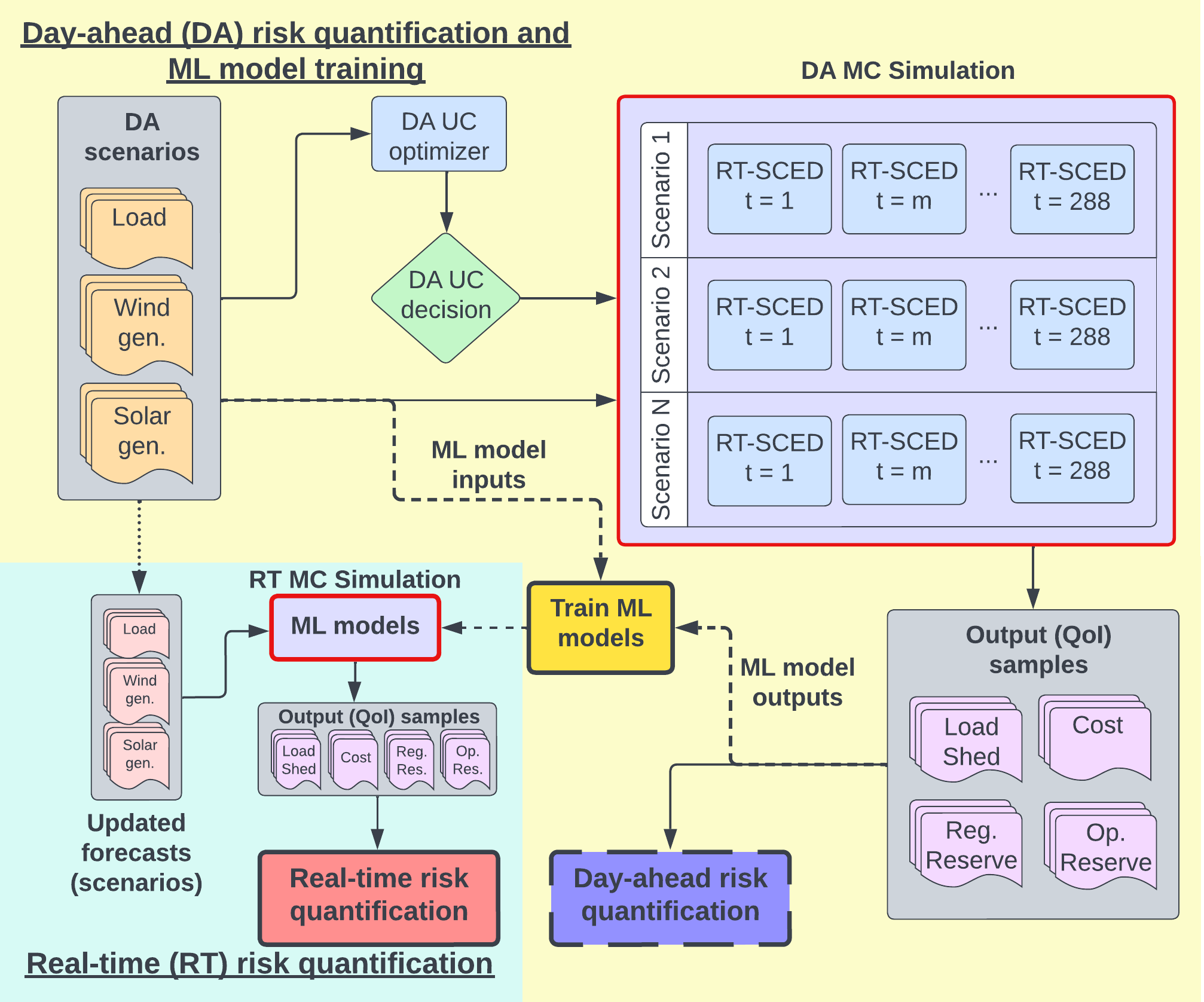}
    \caption{The Just-In-Time Learning Framework for Real-Time Risk Assessment of Power Grids.}
    \label{fig:RAoverview}
\end{figure} 

To reduce the computational requirements of real-time risk assessment, this paper proposes a Just-In-Time Risk Assessment Learning Framework (JITRALF) depicted in Figure \ref{fig:RAoverview}.
The red box in the figure denotes the initial DA risk assessment, which uses the optimization-based methodology outlined in Algorithm \ref{alg:MC_Risk_Assessment}.
Second, following this DA risk assessment, the JITRALF trains {\em risk surrogates}, i.e., machine-learning models that predict  the QoI values computed in line \ref{alg:MC_risk:MC:QoI} of Algorithm \ref{alg:MC_Risk_Assessment} without solving any optimization problem.
This reduces the time and resources needed to perform each MC simulation in Algorithm \ref{alg:MC_Risk_Assessment}.
Third, throughout the day, these risk surrogates are used to perform real-time risk assessment, using the framework of Algorithm \ref{alg:MC_Risk_Assessment} wherein each sequence of optimizations in lines \ref{alg:MC_risk:MC:init}--\ref{alg:MC_risk:MC:QoI} is replaced with an evaluation of the risk surrogate.
For clarity, the MC samples used in day-ahead (resp., real-time) risk assessment are henceforth referred to as day-ahead (resp., real-time) scenarios; recall that DA and RT scenarios are produced by probabilistic forecasting models of suitable granularity.

It is also important to emphasize a few properties of JITRALF.
First, the training of the risk surrogates follows the \emph{just-in-time} learning paradigm introduced in \cite{Chen2022_JustInTimeSCED}, wherein models are trained after the execution of the DA-FRAC, and training data consists of the simulations executed in the DA risk-assessment.
In particular, a separate risk surrogate model is trained for each hour of the day: since commitment decisions vary hourly, they are implicitly captured by the surrogate model trained for that hour.
Second, rather than learning high-dimensional dispatch decisions as in \cite{Chen2022_JustInTimeSCED}, the risk surrogates considered here learn the lower-dimensional QoIs  directly.
This reduction in output dimension allows for smaller models that are easier and faster to train.
Third, once trained, the risk surrogates produce the estimated QoIs in milliseconds.
This makes it possible to perform real-time risk assessment with reasonable computational resources even when considering large number of scenarios, thereby overcoming the computational barrier of the optimization-based approach described in Section \ref{sec:risk}.

The rest of this section presents JITRALF in detail and discusses the spatial granularity of the risk surrogates, their temporal granularity, their underlying structure, and how they capture risk.

\subsection{Spatial Granularity of Risk Surrogates}

This paper focuses on performing system-level risk assessment for a power grid, i.e., the QoIs are defined at the system level.
Namely, the following QoIs are considered: total operating cost (in \$), total load shed (in MW), total regulating reserves (in MW), and total operating reserves (in MW).
The risk surrogates take, as input features, the zonal demand, zonal wind generation, and zonal solar generation, and output the four QoIs.
Zonal features reflect the fact that system operators typically divide power grids into geographical zones, and routinely conduct adequacy and security studies at the \emph{zonal} level.
Finally, recall that, because separate models are trained for each hour of the day, they implicitly capture hourly commitment decisions. The present methodology and surrogate models are not restricted to system-level QoIs nor zonal-level input features.
Future work will consider more granular risk assessment, e.g., at the zonal or nodal level.

\subsection{Temporal Granularity of Risk Surrogates}

Training a unique risk surrogate for an entire month or season is a significant challenge.
Indeed, such a model would have to capture, not only the variability in weather conditions, which influences consumption and renewable production, but also the combinatorial explosion of generator commitments, outages, and topology changes.
Therefore, this work follows the \emph{Just-In-Time} learning paradigm introduced in \cite{Chen2022_JustInTimeSCED}.
This approach leverages the fact that, while generator dispatches are computed every five minutes \cite{tsao_2018}, generator commitments only vary on an hourly basis, and are almost always known as soon as the day-ahead market is cleared and the subsequent FRAC is executed.
This makes it possible for JITRALF to train a separate risk surrogate for each hour of the day, thereby eliminating any variability in generator commitments.

Each risk surrogate is trained using the data generated by the simulations used for the day-ahead risk assessment, and predicts the QoIs for a specific hour of the day.
Namely, when performing real-time risk assessment throughout the operating day, predictions are made using the risk surrogate trained for that hour.
Finally, recall that the RT scenarios of load and renewable production, used to perform RT risk assessment, are produced by a probabilistic forecasting model that predicts the (joint) distribution of these variables for the hour ahead.

\subsection{The Architecture of Risk Surrogates}

The risk surrogates explored in this work are multivariate regression models that predict multiple (4) QOIs, given zonal demand, wind generation, and solar generation.
This research considers various popular parametric and non-parametric regression models.
Non-parametric models provide accurate prediction for QoIs with mixed-continuous behavior, especially, when a large amount of training data covering a wide range of input states is available.
Some of the QoIs, such as the load shed, have a mixed-continuous form: they take a single value (zero) most of the time during the day and deviate from this single value for a short duration.
Since non-parametric models do not assume any structure, they are better suited to capture this mixed-continuous behavior.
Note that non-parametric models are not capable of extrapolation and provide inaccurate predictions for unseen input states, so it is important to include many possible system states in the training data.
Parametric models, on the other hand, may exhibit a lower extrapolation error, but may have trouble in learning QoIs exhibiting mixed-continuous behavior.

\subsection{The Training of Risk Surrogates}
\label{sec:training} 

Unlike many machine-learning applications, the evaluation of risk surrogates should not be based on their mean prediction accuracy. Instead, the surrogate model's ability to predict risk is determined by its accuracy on \emph{extreme} scenarios (i.e., samples whose output violates some threshold), since such scenarios drive system risk. Without the ability to accurately predict QoIs near and/or beyond  safety limits, these surrogates cannot be successfully employed to estimate risk. As a result, to ensure accurate prediction of extreme system behavior, JITRALF must: a) ensure that adequate training data is available from regions near or beyond acceptable operating limits, and b) utilize specialized loss functions and/or validation metrics that give higher weights to output values near/beyond the safety threshold. This subsection discusses these issues.

\noindent\paragraph{Training Data} To ensure that the training data covers the range of possible values of the stochastic input variables, it is important to generate scenarios from a statistically calibrated and properly validated forecasting model. This is discussed in detail in~\cite{LSTM}. Furthermore,
sufficient training data should be available for input regions where the QoIs are near or beyond acceptable operating limits (i.e., unsafe regions). Without such training data, a risk surrogate may not be able to predict the violation of acceptable operating conditions. The training process should use {\em data augmentation} to train the risk surrogate if only a small number of scenarios  violating operating limits are available for the DA risk assessment. Data augmentation is natural in this setting, since the SCUC, FRAC, and SCED can be used to obtain the QoIs on these scenarios. Obviously these scenarios should only be used to train the risk surrogates and ensure that they have the proper accuracy on these unsafe regions, and not to perform DA risk assessment. In essence, this is similar to importance sampling~\cite{glynn1989importance}, where the sampling process is biased to promote failure, followed by a correction to account for the induced bias.

\noindent\paragraph{Loss Function and Validation Metric} In addition to generating adequate training data from unsafe regions, the prediction accuracy needs to be quantified for these unsafe regions (not just for typical scenarios). Existing loss functions and validation metrics (e.g., MSE, MAE, asymmetrical loss functions) do not capture the greater importance of prediction accuracy in these unsafe regions versus safe regions. Furthermore, some DMs may desire \emph{conservative} predictions, particularly when predicting in these unsafe regions. For these reasons, JITRALF introduces the hazard-aware loss (HAL) metric as a novel loss function. Let $q$ and  $\hat{q}$ be the true value of the QoI and its prediction, and let $\bar{q}$ be a threshold specifying the unsafe region, i.e., a QoI value below $\bar{q}$ indicates an unsafe state.\footnote{A similar loss function can be defined when the unsafe states are defined by values exceeding a given threshold.} The loss function HAL is defined as
\begin{align}
  &L(q, \hat{q}, u_{\rm safe},o_{\rm safe},u_{\rm unsafe},o_{\rm unsafe}) \nonumber \\
  & \! = \! \begin{cases}
    u_{\rm safe} \! \cdot \! {\rm min}(\hat{q}-q,0) + o_{\rm safe} \! \cdot \! {\rm max}(\hat{q}-q,0),
     \text{if   } q>\bar{q},\\
    u_{\rm unsafe} \! \cdot \! {\rm min}(\hat{q}-q,0) + o_{\rm unsafe} \! \cdot \! {\rm max}(\hat{q}-q,0) \text{otherwise,}
  \end{cases} \label{eq:HALoss}
\end{align}
where $u_{\rm safe}$ and $o_{\rm safe}$ are penalties for under- and over-estimations in safe regions, and $u_{\rm unsafe}$ and $o_{\rm unsafe}$ are the penalties for under- and over-estimation in unsafe regions. By selecting values of $u_{\rm safe}$, $o_{\rm safe}$ $u_{\rm unsafe}$, and $o_{\rm unsafe}$, a DM can express the preference for under- and over-estimation in both safe and unsafe conditions.

\section{Case Study}
\label{sec:numex}


This section demonstrates JITRALF on an industry-size system, i.e., the French transmission grid.
The grid topology is taken from a snapshot of the real system, and contains over 6500 buses, 9000 transmission lines, and about 2000 generators \cite{Chatzos2022_DataDrivenTimeSeries}.
It is divided into 12 zones, each corresponding to an administrative region in France.
Time series for bus-level load and renewable production, as well as generators' economic data, have been reconstructed following the methodology described in \cite{Chatzos2022_DataDrivenTimeSeries}.

\subsection{Experimental Setting}

The MISO optimization pipeline described in Section \ref{sec:risk_assessment:MISO_operations} was executed on data from  March 16th, 2018.
The DA-SCUC and DA-FRAC were used to produce hourly commitments for all generators.
Multiple DA scenarios were generated for load and renewable (wind and solar) production at every bus for the entire day at a 5-minute granularity.
In practice, 99\% of commitments are decided in the DA-SCUC and DA-FRAC \cite{Chen2018_MarketClearingSoftware}. 
Therefore, for ease of presentation, the experiments did not execute a LAC during the day.
This is without loss of generality: JITRALF can be readily adapted to handle LAC studies.
For each DA scenario, a day-long simulation is performed, solving 288 RT-SCED problems sequentially, one every five minutes, to compute energy and reserve dispatches for all (online) generators.
These dispatch decisions yield corresponding samples of the system-wide operating and regulating reserves, total operating costs, and total load shed.
Using these samples of QoIs, a day-ahead risk profile associated with the selected portfolio of committed generators is computed.

The DA forecasts used in DA-FRAC and the DA scenarios used for simulations are obtained as follows.
Day-ahead, hourly forecasts for total load, total wind production and total solar production are obtained from RTE \cite{eco2mix}.
Then, a nearest-neighbor model identifies the 100 prior operating days that display the closest load and renewable output, yielding an initial set of 100 DA scenarios.
A total of $2{,}500$ DA scenarios are obtained by taking convex combinations of the original 100 DA scenarios, whose weights are sampled from a Dirichlet distribution of dimension $100$ and parameter $10^{-2}$.
Note that JITRALF is agnostic about how the DA scenarios are generated, and other probabilistic forecasting models can be readily used.

\subsection{Day-Ahead Risk Assessment}

\begin{figure}[!ht]
    \centering
    \begin{subfigure}[b]{0.35\textwidth}
        \centering
        \includegraphics[width=\textwidth]{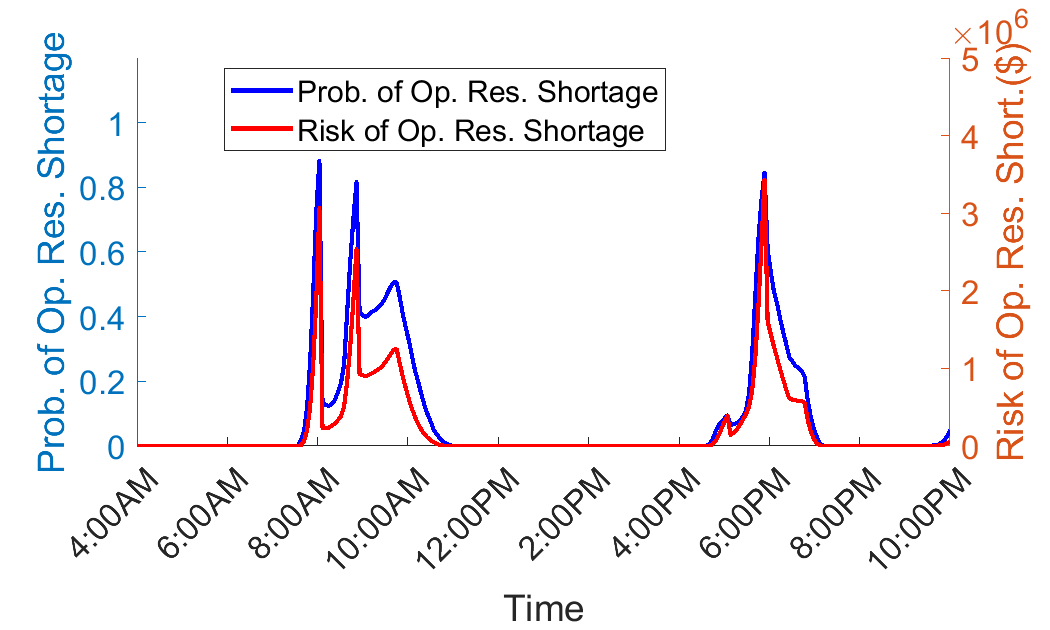}
        \caption{Probability (\textcolor{blue}{\textbf{blue}}, left axis) and Risk (\textcolor{red}{\textbf{red}}, right axis) of Operating Reserve Shortage.}
        \label{fig:DA_op_res}
    \end{subfigure}
    \hfill
    \begin{subfigure}[b]{0.35\textwidth}
        \centering
        \includegraphics[width=\textwidth]{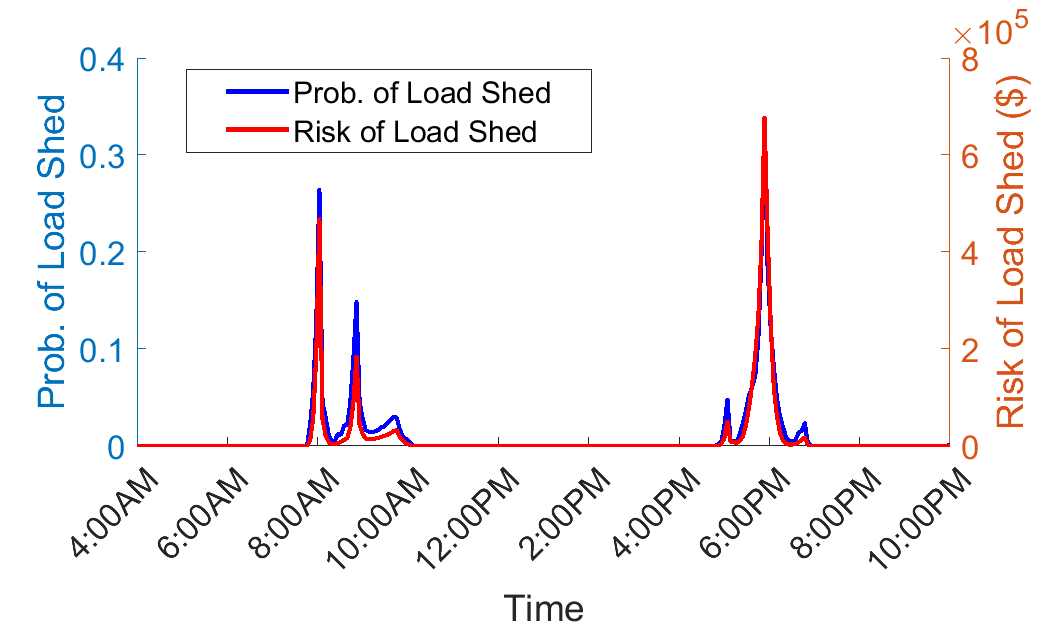}
        \caption{Probability (\textcolor{blue}{\textbf{blue}}, left axis) and Risk (\textcolor{red}{\textbf{red}}, right axis) of Load Shedding.}
        \label{fig:DA_LS}
    \end{subfigure}
    \caption{Day-Ahead Risk Profile.}
    \label{fig:DA_risk}
\end{figure}

The DA probability and expected risk of an operating reserve shortage (operating reserve less than MRR) are shown in Fig.~\ref{fig:DA_op_res}.
The probability and expected risk of a load-shedding event is shown in Fig.~\ref{fig:DA_LS}. In these figures, the probability of the adverse event is shown in blue on the left axis, and the expected risk of the adverse event is shown in red on the right axis. None of the scenarios have regulating reserve values less than the MRR constraints so the probability and expected risk of a regulating reserve shortage (regulating reserve less than MRR) is zero for all time steps. The risk computation assumes that the monetary consequences incurred by adverse events is \$3,500 per MW for load shedding or reserve shortage.
The selected generator portfolio has a high probability and expected risk of insufficient operating reserve and load shedding during morning and evening hours.
A rational grid operator is unlikely to proceed with the selected generator portfolio, given the level of risk. However, the selected portfolio is an interesting setting to demonstrate the utility of JITRALF.

\subsection{Risk Surrogates: Training, Validation and Model Selection}
\label{sec:numex:risk_surrogates}

For the system-level risk, the model inputs are the total and zonal wind generation, solar generation, and load. The model outputs are the operating cost, the regulating reserve capacity, the operating reserve capacity, and the load shedding at the system level. Non-parametric multivariate random forest (RF) and parametric feed-forward neural networks are two candidates for designing the risk surrogates. The multivariate RF regression model was developed using Python's scikit-learn API. Two feed-forward neural networks were trained: one using MAE as the loss function (referred to as NN(MAE)) and one using HAL (referred to as NN(HAL)). Both of these models were built using the Keras API in Python. Each model has an input layer (50 neurons with a ReLU activation function), three hidden layers (30 neurons with a leaky ReLU activation function, 20 neurons with a leaky ReLU activation function, and 10 neurons with a  ReLU activation function), and an output layer (4 neurons with a linear activation function). Both models are trained with the Adam optimizer and early-stopping. For the second model, the weighted loss function was defined with $u_{\rm safe}, o_{\rm safe}, u_{\rm unsafe}$ and $o_{\rm unsafe}$ set to 1.05, 1, 1.2, and 1.1 respectively. For regulating reserves, $\bar{q}$ was set to the MRR value (500 MW), with values below $\bar{q}$ indicating an adverse event. For operating reserve,  $\bar{q}$ was set to the MRR value (2,250 MW), with values below $\bar{q}$ indicating an adverse event. For load shedding, $\bar{q}$ was set to 0 MW, with values above $\bar{q}$ indicating an adverse event. $\bar{q}$ was not defined for the operating cost, since there is no failure mechanism defined for extreme costs. These parameter values were used for all models/metrics. The experiments split the DA scenarios for each hour into a training set and a testing set (70\%-30\% train-test split). The testing error for the RF, NN(MAE) and NN(HAL) models was evaluated using both mean absolute error (MAE) and the HAL metric. The model with the smallest testing error was selected for real-time risk estimation. 

\subsection{Real-Time Risk Assessment}

For a given risk surrogate, JITRALF was evaluated to perform real-time risk assessment. First, three additional DA scenarios are selected that correspond to high-, medium- and low-risk operations. They are selected from an additional set of DA scenarios, not seen during training, so that the cost of load shedding and reserve shortages over the day fall within the highest, middle, and lowest third, respectively. This allows for the evaluation of the risk surrogates under different stress conditions.

\begin{figure}[!ht]
    \centering
    \includegraphics[width=0.45\textwidth]{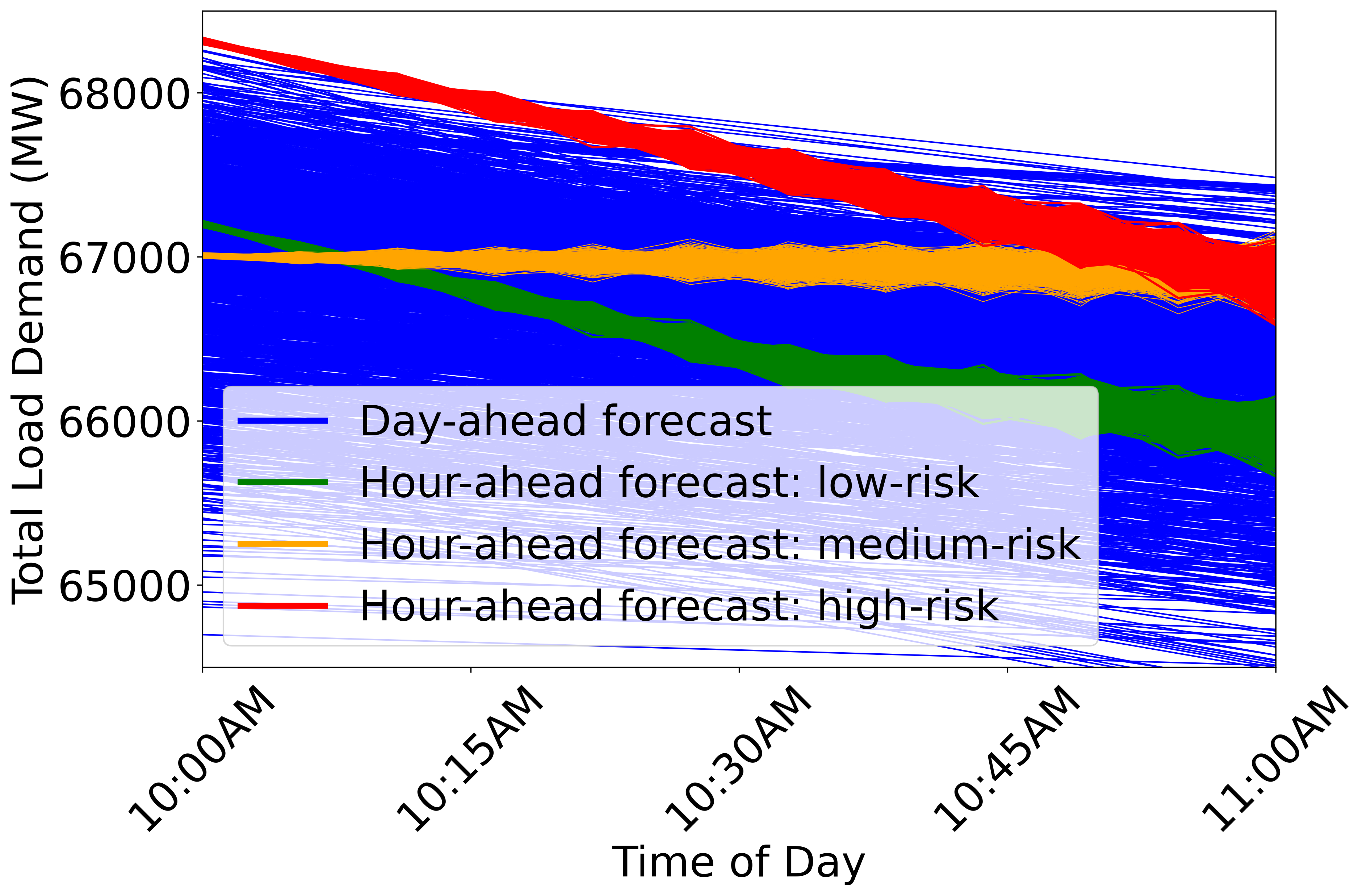}
    \caption{Updated net load demand forecast: three (high-, medium-, and low-) risk scenarios correspond to three days, on which the load demand followed the specified risk regime.}
    \label{fig:updated_forecast}
\end{figure}

Second, for each of the high, medium, and low-risk cases, 288 RT-SCED problems are solved sequentially, one every 5 minutes. In addition, every 15 minutes, a probabilistic forecast for load and renewable production is produced that spans the next hour with a 5-minute granularity; the corresponding samples are referred to as \emph{short-term} (ST) scenarios. For the experiments, 1000 ST scenarios are generated every 15 minutes, by adding geometric Brownian noise to the actual values. The parameters of the Brownian motion are chosen so that the noise's standard deviation after 1 hour represents 2.5\% of the actual values.
Again, the proposed methodology is agnostic to how the samples are generated, and any short-term probabilistic forecasting model may be used.
Figure \ref{fig:updated_forecast} depicts the ST load scenarios for high, medium and low-risk cases against the original DA scenarios used for training.

Third, every 15 minutes, a real-time risk assessment is performed using the 1000 ST scenarios at hand. This risk estimation is performed twice: once using the RT-SCED optimization, wherein 12 RT-SCED models are solved sequentially for each ST scenario, and once using the risk surrogates. The former is referred to as the true risk estimate, and the latter as the JITRALF prediction.
Each simulation takes around 10s with the RT-SCED model, and only a few milliseconds when using the ML surrogate.
Finally, this risk assessment is performed during the morning hours of 5am to 11am, and the evening hours of 4pm to 8pm, which were found in Figure \ref{fig:DA_risk} to be the only time periods to exhibit non-zero risk of an adverse event. The next section evaluates the performance of risk surrogates and  the real-time risk estimation.

\section{Experimental Results}
\label{sec:res}

\subsection{Validation and Selection of Risk Surrogates}


\begin{table}[!ht]
    \centering
    \caption{Normalized MAE for risk surrogate model validation. MAE values are normalized by the mean value for the cost and regulating reserve (\$104408 and 506 MW, respectively), and the mean value in the unsafe region for the operating reserve and load shed (1540MW and 465MW, respectively).}
    \label{table:mdl}
    \resizebox{\columnwidth}{!}{
    \begin{tabular}{lrrrrrr}
    \toprule
        & \multicolumn{3}{c}{Safe} & \multicolumn{3}{c}{Unsafe}\\
        \cmidrule(lr){2-4} \cmidrule(lr){5-7}
    QoI & RF & NN1$^{\dagger}$ & NN2$^{\ddagger}$ & RF & NN1$^{\dagger}$ & NN2$^{\ddagger}$\\
    \midrule
    Cost            & 0.17 & 0.48 & 0.48 & - & - & - \\ 
    Reg. Reserve    & 0.24 & 0.48 & 0.50 & - & - & - \\
    Op. Reserve     & 0.86 & 2.70 & 2.90 & 1.40 & 3.90 & 3.80 \\
    Load Shed       & 0.02 & 0.08 & 0.07 & 5.30 & 35.20 & 28.90 \\
    \bottomrule
    \end{tabular}}\\
    $^{\dagger}$training loss: MAE; $^{\ddagger}$training loss: HAL. All values in \%.
    \label{table:mdl2}
\end{table}

The risk surrogate models are validated using data unseen during the model training process.
As previously discussed, a risk surrogate model must accurately predict QoIs in both safe and unsafe regions.
The experiments  assess the model performance in safe and unsafe regions separately by computing the normalized mean average error (NMAE) in these regions.
These results are reported in Table~\ref{table:mdl2}.
The NMAE metric reported in Table~\ref{table:mdl2} is the MAE value normalized by the average value of the QoI (given in the caption of Table~\ref{table:mdl2}).
The cost and regulating reserve predictions were not evaluated in unsafe regions: a safety limit is difficult to define for operating cost, and the minimum regulating reserve requirements were never violated in training or testing data.
As shown in the Table, all three models produce accurate predictions in the safe regions.
The RF model has the lowest NMAE values, and the two neural networks have similar prediction errors.
However, the distinction between the models is sharper for predictions in the unsafe region.
The RF has substantially lower NMAE values than the neural networks.
The NN trained with the HAL loss function also has lower NMAE value in the unsafe region than the NN trained with an MAE loss function.
This is the expected outcome as the HAL loss is designed to give a higher penalty for prediction error in the unsafe region.
The NN models have high NMAE values when predicting load shed in the unsafe region.
However, the NMAE for RF model for the load shed is only about about 5\%.
Given the substantially better performance of the RF model in both the safe and unsafe regions, this model is selected as the risk surrogate.

\subsection{Real-time Risk Estimation}

\begin{figure*}
     \centering
     \begin{subfigure}[b]{0.3\textwidth}
         \centering
         \includegraphics[width=\textwidth]{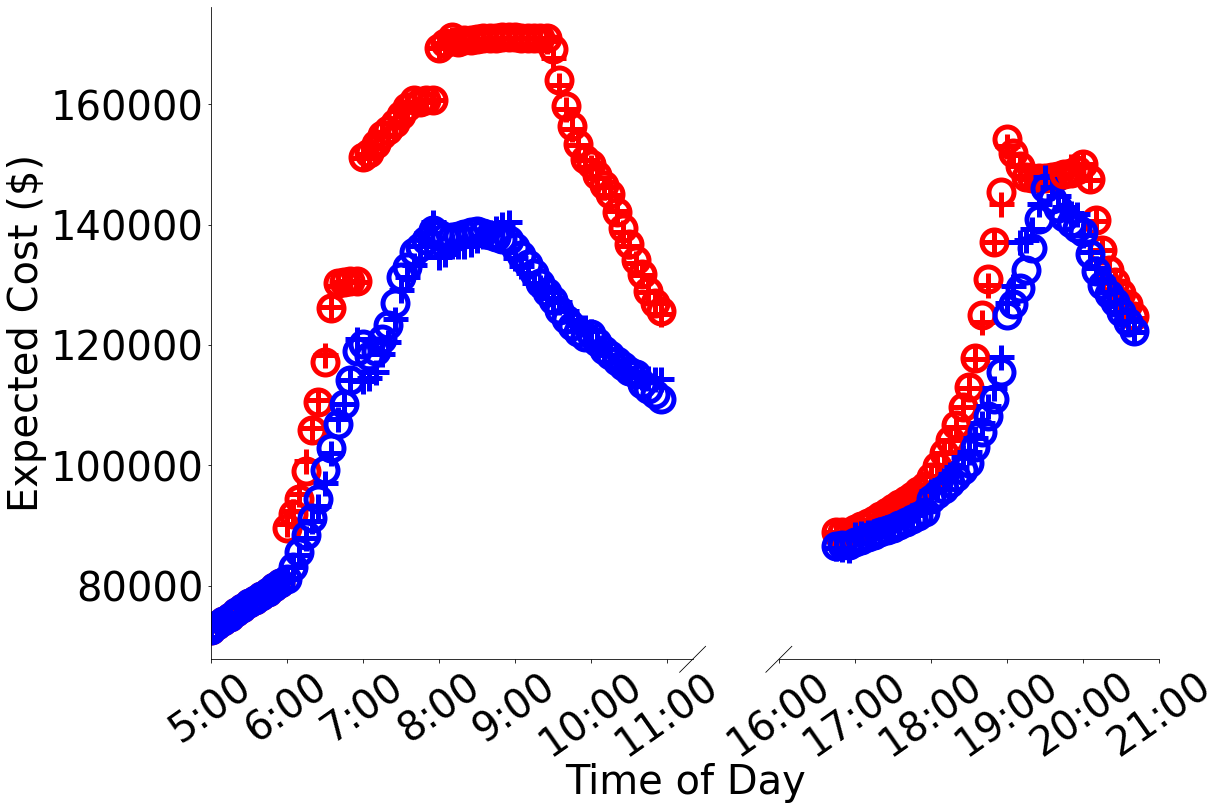}
         \caption{Expected operating cost}
         \label{fig:cost}
     \end{subfigure}
     \begin{subfigure}[b]{0.3\textwidth}
         \centering
         \includegraphics[width=\textwidth]{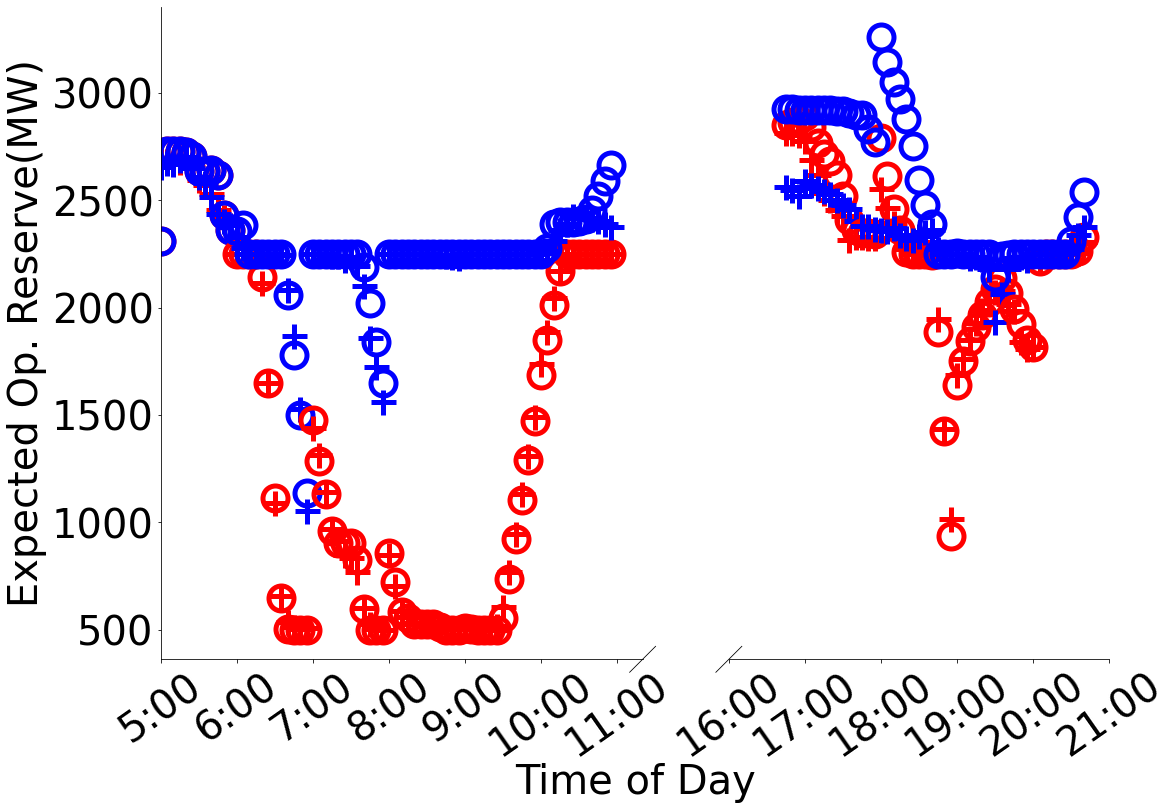}
         \caption{Expected operating reserve capacity}
         \label{fig:op_res}
     \end{subfigure}
     \begin{subfigure}[b]{0.3\textwidth}
         \centering
         \includegraphics[width=\textwidth]{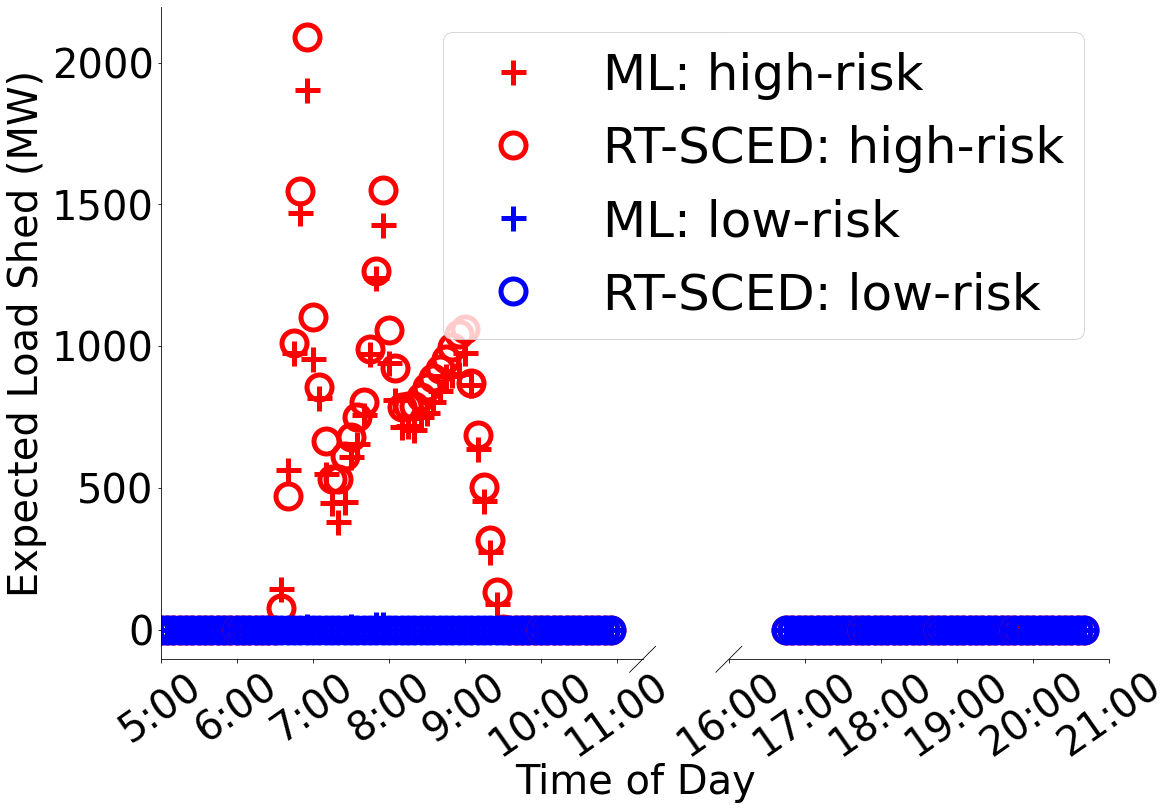}
         \caption{Expected load shed}
         \label{fig:load_shed}
     \end{subfigure}
     
       \caption{Updated QoI predictions}
         \label{fig:res_qois2}
\end{figure*}

A subset of predictions based on SCED and risk surrogates are shown in Fig.~\ref{fig:cost} for the expected cost, in Fig.~\ref{fig:op_res} for the operating reserve capacity, and in Fig.~\ref{fig:load_shed} for load shedding. For clarity, the figures only show the high- and low-risk test cases and predictions obtained at the top of each hour. As shown in these figures, the risk surrogate gives accurate predictions for all QoIs for both cases. While accurate predictions for all QoIs are desirable, only the prediction accuracy in unsafe states impacts the subsequent predictions of the risk of an adverse event. As shown in Fig.~\ref{fig:op_res}, the risk surrogate exhibits some errors in predicting the operating reserve capacity when the true value is above the MRR. It has very accurate predictions when the true value is at or below the MRR. While the errors in safe regions are not desirable, they have no impact on the subsequent risk prediction. This superior predictive performance in unsafe regions will lead to superior accuracy in predicting risk. Note that the HAL validation metric is formulated to mimic this higher weight given to errors in the unsafe region. 

\begin{figure}[!ht]
    \centering
    \begin{subfigure}[b]{0.3\textwidth}
        \centering
        \includegraphics[width=\textwidth]{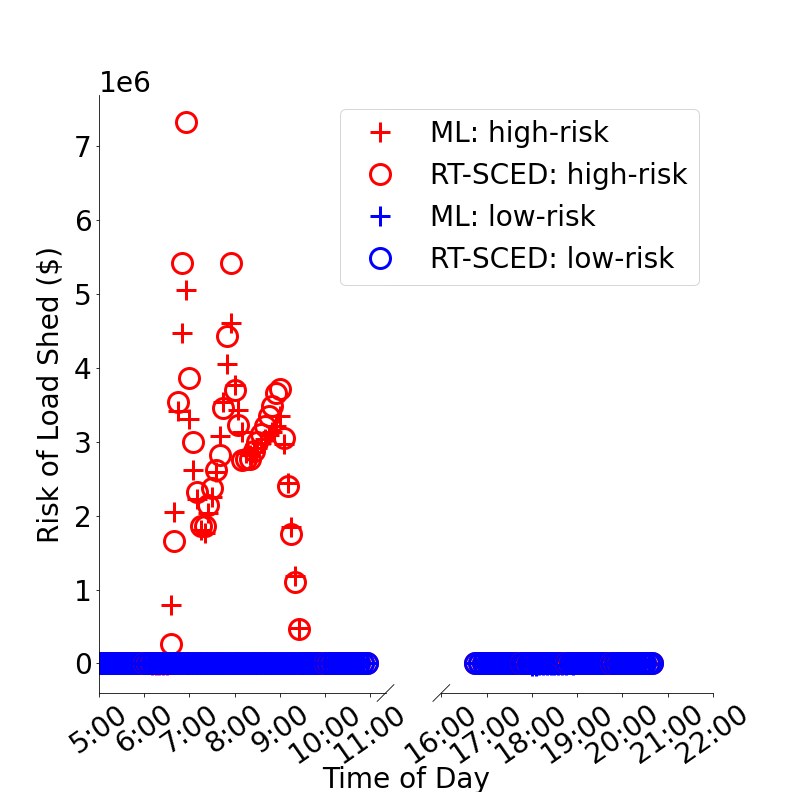}
        \caption{Updated prediction for the risk of load shed}
        \label{fig:ls_risk}
    \end{subfigure}
    \hfill
    \begin{subfigure}[b]{0.3\textwidth}
        \centering
        \includegraphics[width=\textwidth]{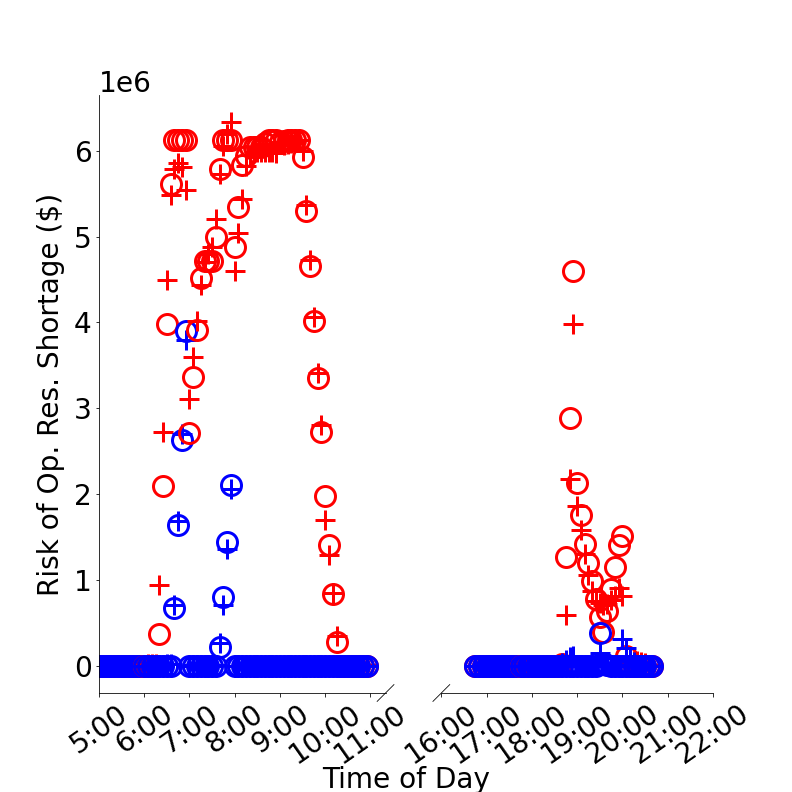}
        \caption{Updated prediction for the risk of operating reserve shortage}
        \label{fig:op_res_risk}
    \end{subfigure}
    \caption{Updated risk profile: load shed and operating reserve shortage}
    \label{fig:res_profiles_op}
\end{figure}

The next results concern the updated predictions for the expected risk of adverse events using the samples generated by the RT-SCED optimizer and the risk surrogates.
Fig.~\ref{fig:ls_risk} reports the predicted expected risk of load shedding and Fig.~\ref{fig:ls_risk} shows the operating reserve shortage.
The validation metrics (MAE and HAL) for risk estimation are shown in Table~\ref{table:risk}. Note that  $\bar{q}$ was set to zero and an event with non-zero risk is considered as an adverse event for computing these metrics.
These results indicate good prediction accuracy for all three cases.
The risk predictions for the low- and medium-risk cases are accurate or conservative, with the risk surrogate predicting a small, non-zero risk of load shedding when the true risk is zero. 
For the high-risk case, the surrogate prediction for the risk of load shedding is slightly lower than the true risk.
This underestimation is not desirable; however, the magnitude of underestimation is small and the surrogate prediction correctly identifies risky conditions.
This model can be used to inform grid operators regarding the relative magnitude of intervention required to mitigate the risk.

\begin{table}[!ht]
    \centering
    \caption{Error ($\times 10^4 \$$) in real-time prediction of risk of adverse event (MAE/HAL). The corresponding mean values of risk are also given for comparison with error values. Accurate real-time risk prediction is possible for credible (non-zero) risk scenarios.}    
    \begin{tabular}{|l c c c|} 
        \hline
        QoI & High-risk  & Medium-risk & Low-risk \\ 
        &  scenario &  scenario &  scenario\\[0.5ex] 
        \hline
        RF &&&\\
        Reg. Reserve  & 0/0 & 0/0 & 0/0 \\
        Op. Reserve & 3.8/4.0 & 1.9/1.9 & 4.2/4.3 \\
        Load Shed  & 8.1/9.7 & 1.4/1.6 & 0.2/0.2 \\
        \hline
        NN (MAE) &&&\\
        Reg. Reserve & 0.12/0.12 & 0.06/0.06 & 0.02/0.02 \\
        Op. Reserve & 9.0/10.4 & 3.0/3.6  & 4.5/5.0 \\
        Load Shed  & 18.1/21.6 & 1.8/2.1 &  0.32/0.33\\
        \hline
        NN (HAL) &&&\\
        Reg. Reserve & 0.28/0.28 & 0.02/0.02 & 0/0 \\
        Op. Reserve & 10.9/12.6 & 2.7/3.1 & 1.6/1.7 \\
        Load Shed  & 8.3/9.8 & 1.3/1.5 & 0.32/0.33 \\ 
        \hline
        \multicolumn{4}{c}{Mean risk (for comparison)}\\
        Reg. Reserve & 0 & 0 & 0 \\
        Op. Reserve & 202 & 58.6 &  10.9  \\
        Load Shed  & 83 & 14.9 & 0 \\ [1ex] 
        \hline
    \end{tabular}
    \label{table:risk}

\end{table}

This example clearly illustrates how cost savings could be achieved by performing real-time risk estimation during daily grid operation.
For the low-risk case, the operator needs to commit only a small amount of additional generating capacity to mitigate the risk.
The high-risk case requires a large amount of additional generating capacity, and the medium-risk scenario requires a moderate amount of additional generating capacity in both the morning and afternoon hours.
If only the day-ahead risk estimate is available, then all three cases are possible and a risk-averse operator may be compelled to commit resources to safeguard the system against the worst (high-risk) case.
If an updated, real-time risk estimate is available, then the operator can remain risk-neutral for a longer time and commit resources only for the risk regime about to be realized on the day of operation.
Thus, by leveraging the real-time risk assessment, grid operators should be able to operate at a lower cost (and potentially lower emission), while maintaining grid reliability and mitigating risk. 

\section{Conclusions}
\label{sec:conclusion}

This paper developed ML-based surrogates for the RT-SCED optimizer to enable real-time risk assessment for commercial power grids.
It discussed key factors that must be considered when training the surrogate models for MC simulation-based risk assessment: 1) training data from unsafe regions, and 2) sufficiently accurate model predictions in unsafe regions.
The paper proposed a just-in-time learning approach to tackle the combinatorial complexity in learning ML models for large scale power systems. Indeed, the proposed approach trains an ML model for each hour of the day to account for changes in grid topology.
To incorporate the decision-maker’s risk-aversion in model training and validation, the paper also proposed a novel loss function called the hazard-aware loss.
This metric captures decision makers' asymmetric aversion to over- and under-estimating risk, particularly in or near unacceptable operating regimes.
The paper used this hazard-aware loss function in evaluating the model performance during validation.
Experimental results demonstrated the real-time risk estimation process on a large industrial system, for three updated forecast cases: high-risk, medium-risk, and low-risk.
Using the real-time risk estimation capability, grid operators can tailor their risk-mitigation intervention based on credible risk-estimates obtained only a few hours in advance, using the most up-to-date information.
They can commit additional resources only after gaining sufficient confidence regarding the occurrence and the magnitude of adverse events.
It is worth emphasizing that various concepts discussed in this work could be used to perform real-time risk assessment of other complex, dynamic systems such as distribution and transportation networks.

The current approach assumes that the real-time generator schedule and grid topology matches the day-ahead schedule and grid topology.
Future research is needed to account for changes to the generator schedule (due to generator failure or operator decisions to dispatch additional generators) and transmission line failures. More research is also needed to improve the extrapolation capability of the ML-based surrogate model and to build surrogates at finer spatial granularity. 

\section{Acknowledgements}

This work was partly funded by the Advanced Research Projects Agency-Energy (ARPA-E) (PI: Prof. Pascal van Hentenryck). 

\bibliographystyle{IEEEtran}
\bibliography{main}

\end{document}